\documentclass[aps,prd,twocolumn,showpacs,preprintnumbers,amsmath,amssymb, longbibliography]{revtex4-1}

\usepackage{bbm}
\usepackage{bm}
\usepackage{color}
\usepackage{dcolumn}
\usepackage{epstopdf}
\usepackage{graphicx}
\usepackage[colorlinks, citecolor=blue]{hyperref}
\usepackage{lipsum}
\usepackage{float}

\newcommand{\beq}{ \begin{equation} }
\newcommand{\eeq}{ \end{equation} }

\newcommand{\e}{\varepsilon}

\newcommand{\down}{\downarrow}

\newcommand{\Op}{\mathcal{O}}

\begin{document}

\title{Time-dependent spintronic anisotropy in magnetic molecules}

\author{Kacper Wrze\'sniewski}
\email{wrzesniewski@amu.edu.pl}

\author{Ireneusz Weymann}
\affiliation{Faculty of Physics, Adam Mickiewicz University, ul. Uniwersytetu Pozna\'nskiego 2, 61-614 Pozna{\'n}, Poland}

\date{\today}


\begin{abstract}
We theoretically study the quench dynamics of
induced anisotropy of a large-spin magnetic molecule
coupled to spin-polarized ferromagnetic leads.
The real-time evolution is calculated by means of the time-dependent density-matrix numerical
renormalization group method implemented within the matrix product states framework,
which takes into account all correlations in very accurate manner.
We determine the system's response to a quench in the spin-dependent coupling to
ferromagnetic leads. In particular, we focus on the transient dynamics associated
with crossing from the weak to the strong coupling regime, where the Kondo correlations
become important. The dynamics is examined by calculating the time-dependent expectation values
of the spin-quadrupole moment and the associated spin operators.
We identify the relevant time scales describing the quench dynamics and
determine the influence of the molecule's effective exchange coupling
and leads spin-polarization on the dynamical behavior of the system.
Furthermore, the generalization of our predictions for
large values of molecule's spin is considered.
Finally, we analyze the effect of finite temperature and show
that it gives rise to a reduction of magnetic anisotropy
by strong suppression of the time-dependent
spin-quadrupole moment due to thermal fluctuations.
\end{abstract}

\maketitle

\section{Introduction}

Molecular magnetism is a rapidly developing area of theoretical and experimental research,
providing concepts for novel applications in spintronic devices and quantum technologies \cite{Boca1999, Tejada2001Jun, Gatteschi2006, Bogani2008Mar, Lehmann2009, Mannini2009Feb, Vincent2012Aug, Jacobson2015Oct, Sessoli2017Aug, Najafi2019Dec}.
Single-molecule magnets (SMM), in particular those of large spin ($S \geqslant 1$), are especially
appealing due to their unique magnetic characteristics
and a wide perspective of engineering and synthesizing new specimen with sought properties \cite{Gatteschi2006}.
One prominent feature present in magnetic molecular systems is the
uniaxial magnetic anisotropy, which leads to the magnetic bistability and suppression
of spin-reversal processes \cite{Gambardella2003May, Hirjibehedin2007Aug, Gambardella2009Feb, Xu2017Aug, Wysocki2020Mar}.
It is a property of crucial importance for the memory storage and information processing applications.
Additionally, when transverse anisotropy component is considerable,
quantum tunneling of magnetization may occur
\cite{Chudnovsky1998Aug, Brechin2002Aug, Misiorny2007Apr, Mannini2010Oct}.
The transport properties of magnetic molecules have already been extensively studied
\cite{Kim2004Apr, Timm2006Jun, Misiorny2009Jun, Misiorny2010Jan, Misiorny2012Jul, Misiorny2015Jan, Plominska2016Jul, Plominska2018Apr, Pawlicki2018Aug, Chiesa2019Jun, Gimenez-Santamarina2019Sep, deBruijckere2019May}, including the influence of the Kondo effect \cite{Kondo1964, Glazman1988, Hewson1997, Goldhaber1998,Kouwenhoven1998} in the strong coupling regime
\cite{Madhavan1998Apr, Romeike2006May, Otte2008Sep, Parks2010Jun, Misiorny2011Mar, Misiorny2012Dec, Zalom2019May}.
However, when the physics of SMM systems incorporates spintronics,
new prominent effects are revealed, including switching with spin-polarized currents,
Berry-phase blockade or spintronic anisotropy among many others
\cite{Misiorny2007Apr2, Gonzalez2007Jun, Misiorny2008May, Delgado2010Jan, Misiorny2013Jul, Misiorny2014Dec, Tyagi2018May, Sierda2019Dec, Verlhac2019Nov}.
In fact, the latter effect is of particular interest,
as it allows for generation of magnetic anisotropy in spin-isotropic
molecules \cite{Misiorny2013Oct}.

Moreover, the dynamics of molecular systems,
an important aspect of the on-going research in molecular magnetism,
has recently gained a lot of attention and has been
explored both experimentally \cite{Johansson2016Nov, Katoh2018Jul, Taran2019May}
and theoretically \cite{Roosen2008, Plominska2017Apr, Hammar2017Dec, Plominska2018Jan}.
Thus, broadening further the knowledge and
understanding of transport and dynamical properties
of large-spin molecules is important both because of exciting fundamental aspects
as well as due to possible applications in modern nanoelectronics, spintronics and quantum information.

Motivated by recent progress within this field,
in this paper we investigate the dynamical behavior of a large-spin magnetic molecule attached to spin-polarized leads,
with an emphasis on the buildup of quadrupolar exchange field, referred to as spintronic anisotropy.
Commonly, the intrinsic magnetic anisotropy arises from the spin-orbit interaction.
However, it has been shown that the ferromagnetic proximity effect
\cite{Martinek2003Sep, Martinek2005Sep, Hauptmann2008Mar, Gaass2011Oct}
can generate significant magnetic anisotropy in molecular systems
in form of an effective exchange field \cite{Misiorny2013Oct}.
The advantage of this approach is the possibility to electrically control both
the magnitude of the anisotropy and the spin state of the system.
When the coupling strength to external contacts is varied,
a rapid change in the magnetic properties of the system occurs \cite{Misiorny2013Oct, Wojcik2019Jul}.
In particular, the molecule's quadrupolar moment is significantly reduced,
when the system is tuned from the weak to the strong coupling regime.
This rapid change of the moment is due to the Kondo screening of the molecule's orbital level spin
and is a non-trivial many-body effect resulting from the interplay of magnetism and the Kondo physics.
It was also shown experimentally that quadrupolar interaction in SMM systems
has an important influence on tunneling dynamics \cite{Taran2019Apr}.
In a real setup, tuning between the different coupling regimes
can be achieved by electrically shifting the tunnel barriers with respective gates.

Considering all the above, we focus on the theoretical study of the quench dynamics of the spintronic quadrupole moment
due to the Kondo correlations. In particular, we identify the universal time scale
for the dynamics describing the quench of spintronic anisotropy.
Moreover, we examine the influence of the magnitude of effective exchange coupling,
leads spin-polarization and total spin of the molecule on discussed dynamical effects.
Lastly, we also analyze the influence of finite temperature, showing that in certain range
of temperatures, a strong suppression of magnetic properties is predicted.
In pursuance of the precise analysis of the system's response
to the considered quench in the strong coupling regime,
we resort to the Wilson's numerical renormalization
group (NRG) method \cite{Wilson1975, Bulla2008, NRG_code}.
We use the extended implementation allowing for studying
the time evolution of the system, namely,
the time-dependent numerical renormalization group (tNRG)
\cite{Anders2005, Anders2006, Costi2014generalization, Costi2014, Costi2018}.
This method allows for taking into account
all the correlations in a fully non-perturbative manner
and, thus, generating reliable predictions for the dynamics
of the system under investigation.

This paper is structured as follows.
Section \ref{theoretical framework} consists of
the Hamiltonian description of the considered system,
the overview of the quench protocol and a summary of
the numerical renormalization group method used for
calculations of time-dependent expectation values of local observables.
In Sec. \ref{results} we present the numerical results
and relevant analysis for the quantum quenches in the coupling
strength from the weak to the strong coupling regime.
We also present and discuss the effects of finite temperature on dynamical behavior.
Finally, the work is concluded in Sec. \ref{conclusions}.

\section{Theoretical framework} \label{theoretical framework}
\subsection{Hamiltonian}

The effective spin Hamiltonian of a molecular magnet
expressed only with spin coordinates can be written as
\begin{equation}\label{Eq:HamiltonianEff}
H_{\rm eff}=B S_z + D \mathcal{Q}_{zz}.
\end{equation}
Here, $S_z$ is the $z$-th component of the total spin $S$,
$\mathcal{Q}_{zz} \equiv S^2_z- S(S+1)/3$ is the $z$-th diagonal tensor element
of the spin-quadrupole moment,
$B$ is a dipolar field corresponding to external magnetic field,
$D$ is a quadrupolar field related to intrinsic spin-orbit interaction.
This approximate approach is well-established for convenient description
of the system's spectrum and is often used to interpret the spectroscopic data \cite{Gatteschi2006}.
In our considerations, however, the above-introduced quantities are
generated purely by the spin-dependent coupling to ferromagnetic leads \cite{Martinek2005Sep,Misiorny2013Oct},
which allows for tuning of both $B$ and $D$ by electrical means---the property that
makes this approach advantageous from the application point of view.

\begin{figure}[t]
	\includegraphics[width=1\columnwidth]{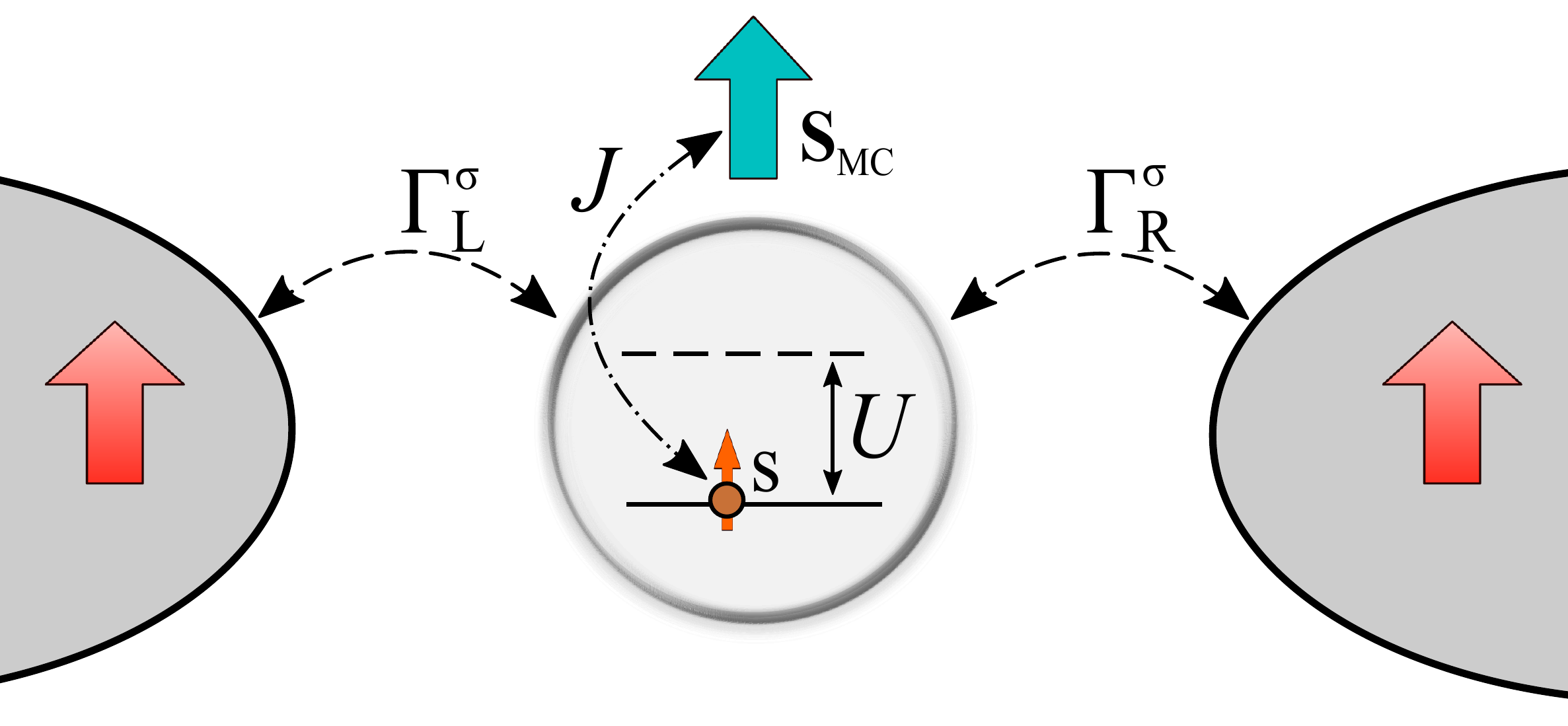}
	\caption{\label{Fig:1}
		Schematic of the considered system.
		A large-spin molecule with a molecular level
		is coupled to external spin-polarized leads
		with the spin-dependent coupling strengths $\Gamma^\sigma_{\rm L}$
		and $\Gamma^\sigma_{\rm R}$, for the left and right lead.
		The molecular level is exchange-coupled
		to magnetic core spin of the molecule with strength $J$.
		Coulomb correlations of the molecule are denoted by $U$.
	}
\end{figure}

In order to analyze the dynamical behavior of the system with tNRG
and, especially, capture all the ferromagnetic-proximity induced effects,
we model the entire system in the following way.
Magnetic molecule is described by a single molecular level,
through which the electronic transport takes place,
which is exchange-coupled to magnetic core of the molecule
specified by the effective spin $\textbf{S}_{\rm MC}$.
Thus, the molecule can be expressed by the Hamiltonian
\begin{equation}\label{Eq:HamiltonianML}
    H_{\rm SMM}=\underbrace{\varepsilon n + U n_{\uparrow}n_{\downarrow}}_{molecular \: level} - J \textbf{S}_{\rm MC} \cdot \textbf{s},
\end{equation}
where the level occupation is expressed as,
$n=n_\uparrow + n_\downarrow=d^\dagger_\uparrow d_\uparrow + d^\dagger_\downarrow d_\downarrow$,
with $d^\dagger_\sigma$($d_\sigma$) being the fermionic creation (annihilation) operator
for an electron with spin $\sigma$. The molecular level energy is denoted by $\e$
and the Coulomb correlations are described by $U$.
We assume ferromagnetic exchange interaction $J>0$
between the spin of the electron on the orbital level $\textbf{s}$
and the magnetic core spin $\textbf{S}_{\rm MC}$.
The total spin is then expressed as $S=\textbf{S}_{\rm MC}+\textbf{s}$.

The molecule is coupled to left and right spin-polarized ferromagnetic leads
\cite{Martinek2003, Choi2004, Martinek2005, Sindel2007}, see Fig. \ref{Fig:1}.
Here, we exploit the correspondence between the coupling to two
leads at equilibrium with the magnetic moments arranged in the parallel configuration
and the coupling to a single ferromagnetic lead.
The equivalence can be shown by carrying out an orthogonal transformation \cite{Glazman1988},
after which the central part of the system couples exclusively to even linear combination
of reservoir's operators with effective coupling strength
$\Gamma^\sigma = \Gamma^\sigma_{\rm L}+\Gamma^\sigma_{\rm R}$
and spin polarization $p$.
Consequently, the leads can be described by an effective reservoir of noninteracting quasiparticles
\begin{equation}\label{Eq:HamiltonianLeads}
  H_{\mathrm{Lead}}=\sum_{\textbf{k}\sigma}\varepsilon_{\textbf{k}\sigma} c^\dagger_{\textbf{k}\sigma} c_{\textbf{k}\sigma},
\end{equation}
where $c^\dagger_{\textbf{k}\sigma}$($c_{\textbf{k}\sigma}$)
is the creation (annihilation) operator of an electron
with momentum $\textbf{k}$, spin $\sigma$ and energy $\varepsilon_{\textbf{k}\sigma}$,
which is given by appropriate linear combination
of electron operators in the left and right leads.
On the other hand,  the spin-dependent coupling is specified by the tunneling term
\begin{equation}\label{Eq:HamiltonianTun}
  H_{\mathrm{Tun}}=\sum_{\textbf{k} \sigma} V_{\sigma} (c^\dagger_{\textbf{k}\sigma}d_{\sigma}  + {\rm H.c.}),
\end{equation}
where $V_{\sigma}$ are the effective tunnel matrix elements,
assumed to be momentum independent.

The spin-dependent coupling between the molecule and the effective lead
is expressed as, $\Gamma^\sigma = \pi \rho^\sigma |V_\sigma|^2$, with
$\rho^\sigma$ being the spin-dependent density of states of ferromagnetic electrodes.
By introducing the spin polarization of the leads $p$,
the coupling strength can be written in the following manner,
$\Gamma^{\uparrow(\downarrow)}=\Gamma(1\pm p)$,
with $\Gamma^{\uparrow(\down)}$ denoting the coupling
to the spin-up (spin-down) electron band of the ferromagnetic reservoir
and $\Gamma= (\Gamma^\uparrow + \Gamma^\downarrow)/2$.

Finally, the full Hamiltonian of the considered system reads
\begin{equation}\label{Eq:HamiltonianTotal}
H=H_{\mathrm{SMM}}+H_{\mathrm{Lead}}+H_{\mathrm{Tun}}.
\end{equation}

\subsection{Quench protocol and NRG implementation}

The time-dependent Hamiltonian describing the
dynamics after a quantum quench has the following general form
\begin{equation}\label{Eq:Hamiltonian_TD}
  H(t) = \theta(-t)H_0 + \theta(t)H,
\end{equation}
where the Hamiltonian $H_0$ denotes the initial Hamiltonian of the system.
On the other hand, $H$ is the Hamiltonian
describing the time evolution after the sudden quench at time $t=0$ and
$\theta(t)$ is the Heaviside step function. Both Hamiltonians have a form
outlined in Eq.~(\ref{Eq:HamiltonianTotal}) with appropriate parameters modified according to the
evaluated quench.
The time-dependent expectation value of
a given local operator $\Op(t)$ can be calculated from
\begin{eqnarray}\label{Eq:O}
  O(t) \equiv \langle \Op(t) \rangle = \mathrm{Tr}\left\{e^{-iHt} \rho_0 e^{iHt} \Op\right\}.
\end{eqnarray}
Here, $\rho_0$ is the initial density matrix of the system
described by the Hamiltonian $H_0$.

Let us now briefly discuss the most important aspects concerning
the NRG implementation of the quench calculations \cite{Wilson1975, Bulla2008, NRG_code}.
The essential part of the NRG procedure is
the logarithmic discretization of the conduction band
followed by mapping of the discretized Hamiltonian to
a one-dimensional tight-binding chain called the Wilson chain \cite{Bulla2008}.
This is performed for both Hamiltonians $H$ and $H_0$.
Subsequently, the two Hamiltonians are independently
solved in an iterative fashion using the NRG procedure \cite{NRG_code}.
At each step of iteration, there are states that are used to construct
the state-space of the next iteration and the states that are discarded.
The discarded states are used to create the full many-body eigenbases \cite{Anders2005}
\begin{equation} \label{eq:completeness}
\sum_{nse}|nse\rangle^{\!D}_{0} \,{}^{D}_{\,0}\!\langle nse| \!=\! \mathbbm{1} \;\;\;\,
 {\rm and}
 \,\;\;\; \sum_{nse}|nse\rangle^{\!D} \,{}^D \!\langle nse| \!=\! \mathbbm{1},
\end{equation}
of both Hamiltonians, $H_0$ and $H$, respectively and to
construct the full density matrix $\rho_0$ at temperature $T \equiv 1/\beta $ \cite{Andreas_broadening2007}
\begin{equation}
\rho_0=\sum_{nse}\frac{e^{-\beta E_{0ns}^D}}{Z} |nse\rangle^{\!D}_{0} \,{}^{D}_{\,0}\!\langle nse| ,
\end{equation}
where
\begin{equation}
Z\equiv\sum_{nse} e^{-\beta E_{0ns}^D}
\end{equation}
is the partition function. Here, $s$ denotes a state at Wilson site $n$,
while $e$ corresponds to an environmental state describing the rest of the chain.

The time-dependent expectation value $\langle \Op(t) \rangle$ of an operator $\Op$
can be conveniently evaluated in the frequency space
and then Fourier-transformed to the time domain \cite{Andreas2012}.
The frequency-dependent expectation value $\langle \Op (\omega) \rangle$
of a local operator $\Op$ expressed in the corresponding
eigenstates of the two Hamiltonians is given by
\cite{Wrzesniewski2019Jul}
\begin{eqnarray}\label{Eq:Ow}
    \langle \Op(\omega) \rangle &=&\!\!
   \sum_{n}^{ XX'\neq KK}\sum_{n'} \sum_{ss'e}  {}^{X}\! \langle nse|w_{n'} \rho_{0n'}| ns'e\rangle^{\! X'} \nonumber\\
   &&\times {}^{ X'}\! \langle ns'e|\Op|nse\rangle^{\! X} \; \delta(\omega + E_{ns}^{X} - E_{ns'}^{X'}).
\end{eqnarray}
where $X = K (X = D)$ denotes a kept (discarded) state.
Here, $\rho_{0n'}$ is the contribution of the density matrix
coming from iteration $n'$ and $w_{n'}$ is the corresponding weight.

We also use NRG to determine the linear-response conductance
between the two ferromagnetic leads from the following formula \cite{Meir1992Apr}
\begin{equation} \label{eq:conductance}
G=\frac{e^2}{h}\pi \Gamma \!\! \int \!\! d \omega  \left ( \!\! -\frac{\partial f}{\partial \omega} \right )
\! [(1+p)A_\uparrow(\omega) + (1-p)A_\downarrow(\omega)],
\end{equation}
where $A_\sigma(\omega)$ is the molecular level's spectral function,
defined as $A_\sigma(\omega)=-(1/\pi){\rm Im} \langle \! \langle d_\sigma|d^\dagger_\sigma \rangle \! \rangle^R_\omega$,
with $\langle \! \langle d_\sigma|d^\dagger_\sigma \rangle \! \rangle^R_\omega$
being the Fourier transform of the retarded Green's function
$\langle \! \langle d_\sigma|d^\dagger_\sigma \rangle \! \rangle^R_t=-i \theta(t) \langle \{d_\sigma(t),d_\sigma^\dagger(0)\}\rangle$.

For the NRG calculations we used the discretization
parameter $2 \leqslant \Lambda \leqslant 3$,
set the length of the Wilson chain to be $N=80$
and kept at least $N_K=4000$ energetically lowest-lying states at each iteration.
In order to suppress the band discretization effects,
we also used the Oliveira's $z$-averaging \cite{Oliveira1990}
by performing calculations for $N_z=4$ different discretizations.
More details and technicalities concerning the implementation
of calculations can be found in Ref. [\onlinecite{Wrzesniewski2019Jul}].


\section{Results and discussion} \label{results}


\subsection{Static properties of the molecule}

\begin{figure}[t]
	\includegraphics[width=1\columnwidth]{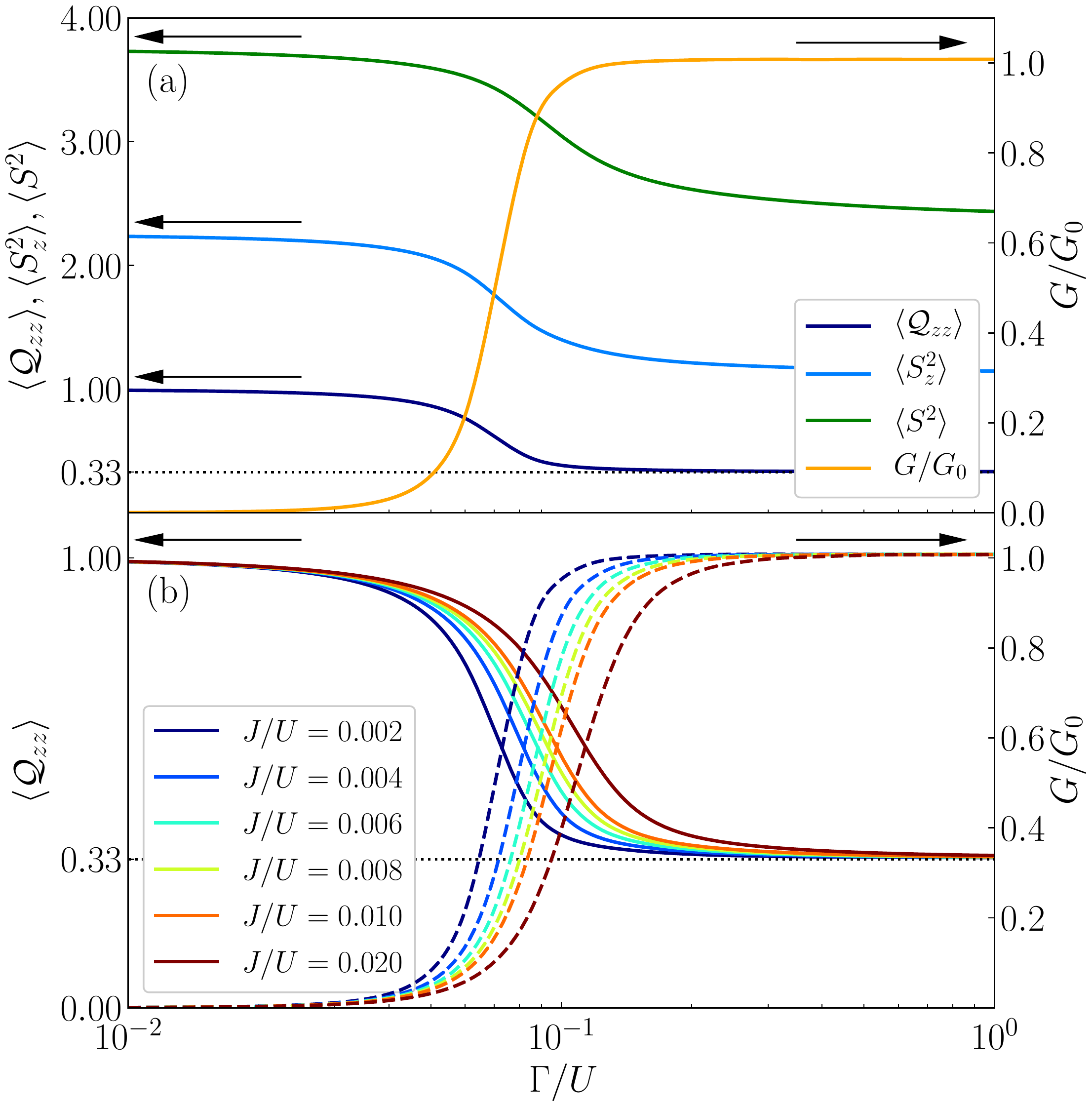}
	\caption{\label{Fig:2}
		(a)  The spin-quadrupole moment $\langle \mathcal{Q}_{zz} \rangle$
		expectation values of the corresponding spin operators, $\langle S^2 \rangle$ and $\langle S^2_z \rangle$,
		and normalized linear-response conductance $G$, with $G_0 = 2e^2/h$,
		for $S=3/2$ spin molecule as a function of the coupling strength $\Gamma$.
		The parameters are: $U=1/2$, $\varepsilon/U=-1/2$, $J/U=2 \cdot 10^{-3}$,
		in units of band halfwidth $W\equiv1$, $p=0.5$ and temperature $T/U \sim 10^{-18}$.
		(b) The spin-quadrupole moment $\langle \mathcal{Q}_{zz} \rangle$ (solid lines) and
		normalized linear-response conductance $G$ (dashed lines)
		plotted versus $\Gamma$ for different values of exchange coupling $J$.
    }
\end{figure}

In order to obtain a better understanding of the magnetic correlations
present in the considered system,
let us first examine the static properties.
As the main focus is put on the quadrupolar field,
we tune the system to the particle-hole symmetry point by setting
the energy of the orbital level to $\varepsilon=-U/2$.
As a result, the dipolar field vanishes \cite{Martinek2003Sep} and
only quadrupolar term is present in the system [second term of Eq.~(\ref{Eq:HamiltonianEff})].
This approach allows us to precisely describe the generated
uniaxial anisotropy, quantified by the amplitude $D$, see Eq. (\ref{Eq:HamiltonianEff}).

In Fig. \ref{Fig:2}(a) we present the spin-quadrupole moment $\langle \mathcal{Q}_{zz} \rangle$
and expectation values of the corresponding spin operators, $\langle S^2 \rangle$ and $\langle S^2_z \rangle$,
as the coupling strength $\Gamma$ is varied.
The general behavior of spin-quadrupole moment (dark-blue line)
is that in the weak-coupling regime ($\Gamma/U \lesssim 10^{-1}$),
it acquires the value $\langle \mathcal{Q}_{zz} \rangle = S(2S-1)/3=1$,
while in the strong-coupling regime ($\Gamma/U \gtrsim 10^{-1}$)
this value is strongly reduced to $\langle \mathcal{Q}_{zz} \rangle = \textbf{S}_{\rm MC}(2\textbf{S}_{\rm MC}-1)/3 = 1/3$.
The suppression of the moment in the strong-coupling regime
is due to the presence of the Kondo correlations.
The Kondo effect is exposed in the conductance dependence (yellow line),
as it saturates to unitary value $G/G_0=1$ in the strong coupling regime.
In consequence, the spin of the molecular level is screened,
and the total spin of the molecule is reduced from $S=3/2$ to $S=\textbf{S}_{\rm MC}=1$,
leading eventually to $\langle \mathcal{Q}_{zz} \rangle = 1/3$.
To clearly show how the expectation values of spin operators
influence the value of the spin-quadrupole moment,
we also plot $\langle S^2 \rangle$ and $\langle S^2_z \rangle$
as a function of the coupling strength.
It is noteworthy that both quantities in the Kondo regime do not achieve the value,
that would be expected when electron on the orbital level was fully screened by the Kondo correlations.
The expected values only approach this limit,
i. e. $\langle S^2 \rangle \rightarrow 2$, $\langle S^2_z \rangle \rightarrow 1$,
however, the resulting value of the spin-quadrupole moment is indeed
$\langle \mathcal{Q}_{zz} \rangle = 1/3$.

In Fig. \ref{Fig:2}(b) we present the spin-quadrupole moment
$\langle \mathcal{Q}_{zz} \rangle$ (solid lines)
and the linear-response conductance $G$ (dashed lines)
as a function of $\Gamma$ for different values of the exchange coupling $J$.
As evident, when the magnitude of $J$ is varied,
the spin-quadrupole moment values in the weak and strong coupling regimes are
conserved, however, the increase of the exchange coupling
extends the transitional range of coupling strength
where the crossover between the weak and Kondo regimes develops.
Although the considered model is an effective one,
we expect that in the case of molecules with strong exchange couplings
between localized spins and those of itinerant electrons,
the quench in the coupling strength needs to be superior
than in the case of systems with small magnitudes of the exchange couplings.
The role of magnitude of $J$ on the dynamics of the spin-quadrupole moment
is discussed in more detail in Sec. \ref{sec:J}.

\subsection{Dynamics of quadrupole moment}

The important alteration in the SMM's magnetic properties
is when the coupling strength $\Gamma$ is switched from the weak coupling regime,
where the spin-quadrupole moment is saturated acquiring
$\langle \mathcal{Q}_{zz} \rangle = S(2S-1)/3=1$,
to the strong coupling regime. In the latter case,
the Kondo correlations are present and the moment is reduced to
$\langle \mathcal{Q}_{zz} \rangle = \textbf{S}_{\rm MC}(2\textbf{S}_{\rm MC}-1)/3 = 1/3$.

\begin{figure}[t]
	\includegraphics[width=1\columnwidth]{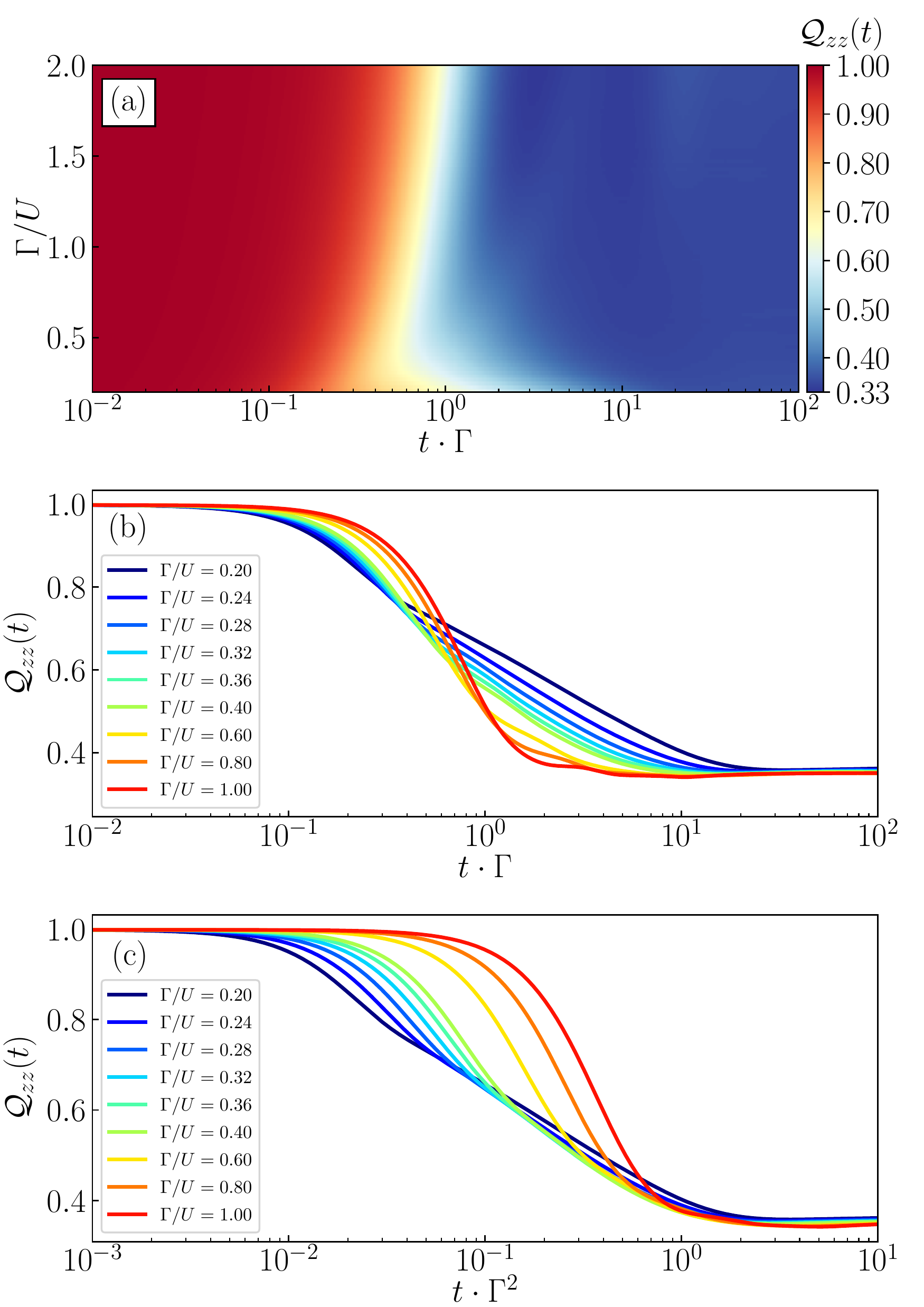}
	\caption{\label{Fig:3}
		The time dependence of the spin-quadrupole moment $\mathcal{Q}_{zz}(t)$
		after quench in the coupling strength from
		$\Gamma_0/U=0.002$ to different values of the  final coupling strength.
		(a) presents the density plot of  $\mathcal{Q}_{zz}(t)$,
		while (b) and (c) show  $\mathcal{Q}_{zz}(t)$
		as a function of $t \cdot \Gamma$ and $t \cdot \Gamma^2$, respectively.
		The other parameters are the same as in Fig. \ref{Fig:2}.
	}
\end{figure}

In order to study the dynamics of this effect,
we perform a quantum quench of the initial Hamiltonian
$H_0$ in the coupling strength from $\Gamma_0/U=0.002$
to various final values of $\Gamma/U$ in the range
of strong coupling where the Kondo effect is  well-established.
In Fig. \ref{Fig:3}(a) we display the resulting time-dependence
of the spin-quadrupole moment $\mathcal{Q}_{zz}(t)$.
First of all, independently of $\Gamma$, both initial $\mathcal{Q}_{zz}(t=0)$
and final $\mathcal{Q}_{zz}(t \rightarrow \infty)$ are in agreement with thermal expectation values.
As can be seen, the long-term value is also
independent of the coupling strength $\Gamma$ as long as the final coupling
strength is within the strong coupling regime, here $\Gamma/U \gtrsim 0.1$, indicated by the unitary
conductance, see Fig. \ref{Fig:2}.
However, the value of $\Gamma$ clearly influences the dynamics of the transition.
In the short-time limit, the initial value of $\langle \mathcal{Q}_{zz} \rangle = 1$
holds for at least time $t \cdot \Gamma \approx 10^{-1}$.
The reduction of quadrupole moment
takes place for times $t \cdot \Gamma \gtrsim 10^{-1}$,
and the dependence of this process is influenced by the final coupling strength.
In the range of $0.4 \gtrsim  \Gamma/U \gtrsim  0.2$,
the drop of $\mathcal{Q}_{zz}(t)$ is observable
as early as $ t \cdot \Gamma \approx 2 \cdot 10^{-1}$
and at middle rate approaches the long-time limit for
times $ t \cdot \Gamma \gtrsim 10^{1}$. However, when the quench
is evaluated for higher values of the final coupling
strength, the time-dependence is more rapid.
On one side, $\mathcal{Q}_{zz}(t)$ starts to drop at later times,
$t \cdot \Gamma \approx  5 \cdot 10^{1}$ for $\Gamma/U =2$,
but on the other side, it achieves long-time limit significantly faster,
i. e. $ \mathcal{Q}_{zz}(t=2/\Gamma)  \approx   \mathcal{Q}_{zz}(t \rightarrow \infty) $
also for $\Gamma/U =2$.
Therefore, by tuning the value of the final coupling strength $\Gamma$,
the timescale of the considered transition can be varied by an order of magnitude.

To clearly display the above behavior, in Fig. \ref{Fig:3}(b) we show the
time evolution of quadrupolar moment for different values of $\Gamma$
chosen from the range, where the dynamics is most interesting.
Finally, in Fig. \ref{Fig:3}(c) we show the same results in
the form of several curves on a rescaled time axis in order to clearly indicate
that the relaxation is universally governed by $t\propto 1/\Gamma^2$.
The long-time limit is achieved at times $t \approx 2/\Gamma^2$,
when one can see that all the curves converge.

\subsection{Role of molecule's exchange coupling}\label{sec:J}

\begin{figure}[t]
  \includegraphics[width=1\columnwidth]{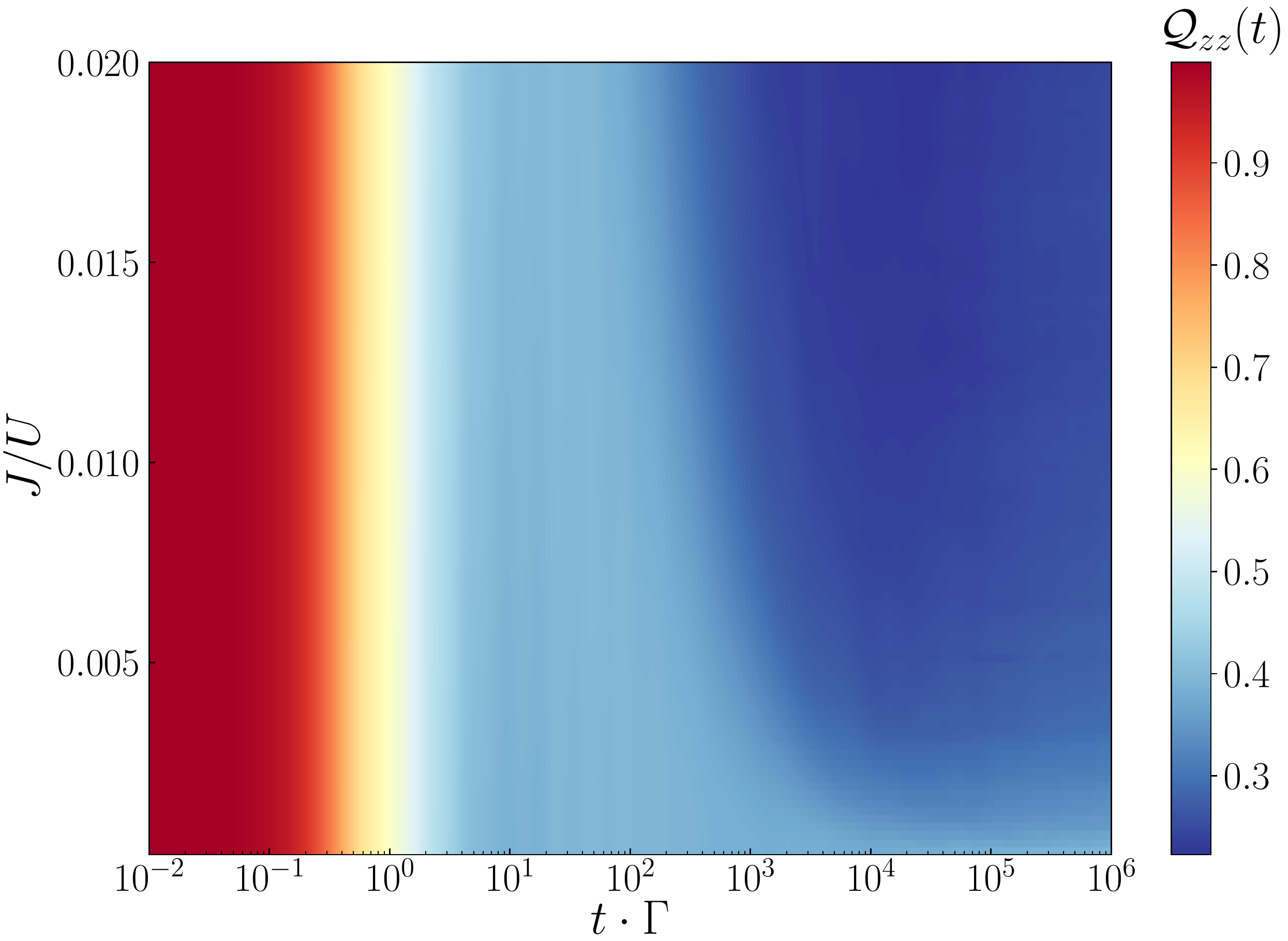}
  \caption{\label{Fig:4}
  The spin-quadrupole moment $\mathcal{Q}_{zz}(t)$
  after the quench from the weak
  to the strong coupling regime ($\Gamma_0/U=0.002$ and $\Gamma/U = 0.4$)
  as a function of time and exchange interaction $J/U$.
  The other parameters are the same as in Fig. \ref{Fig:2}.
}
\end{figure}

An important impact on the magnetic and transport properties of SMM systems
has the magnitude of exchange coupling $J$ between the effective magnetic core spin
$\textbf{S}_{\rm MC}$ and the spin $\textbf{s}$ of electrons occupying the orbital level.
In Fig. \ref{Fig:4} we show the time-dependent spin-quadrupole moment,
considering similar quench as in previous section,
with $\Gamma_0/U=0.002$ and $\Gamma/U = 0.4$, as a function of exchange coupling $J$.

It can be clearly seen that the magnitude of $J$ does not have
a substantial influence on the short time evolution
of spin-quadrupole moment for elapsed time
up to ${t \cdot \Gamma \approx 10^2}$. For $J/U \lesssim 0.002$,
the long-time limit $\langle \mathcal{Q}_{zz} \rangle = 1/3$ is achieved
already around $t \cdot \Gamma \approx 10^1$
and there is no further dynamics as the time elapses.
However, when the exchange coupling is increased above
$J/U \approx 0.002$, further decrease of spin-quadrupole moment takes place at long times,
i.e. for $t \cdot \Gamma \gtrsim 10^2$.
The observed reduction is down to values well below
$\langle \mathcal{Q}_{zz} \rangle = 1/3$, i.e.
$\mathcal{Q}_{zz}(t \rightarrow \infty) \lesssim 0.3$, and
the dynamics of this process is strongly dependent on the magnitude of $J$.
When the time-dependence for larger values of $J$ is evaluated,
the new long-time limit is achieved at earlier times,
while the value of spin-quadrupole moment is accordingly further suppressed.

\begin{figure}[h]
  \includegraphics[width=1\columnwidth]{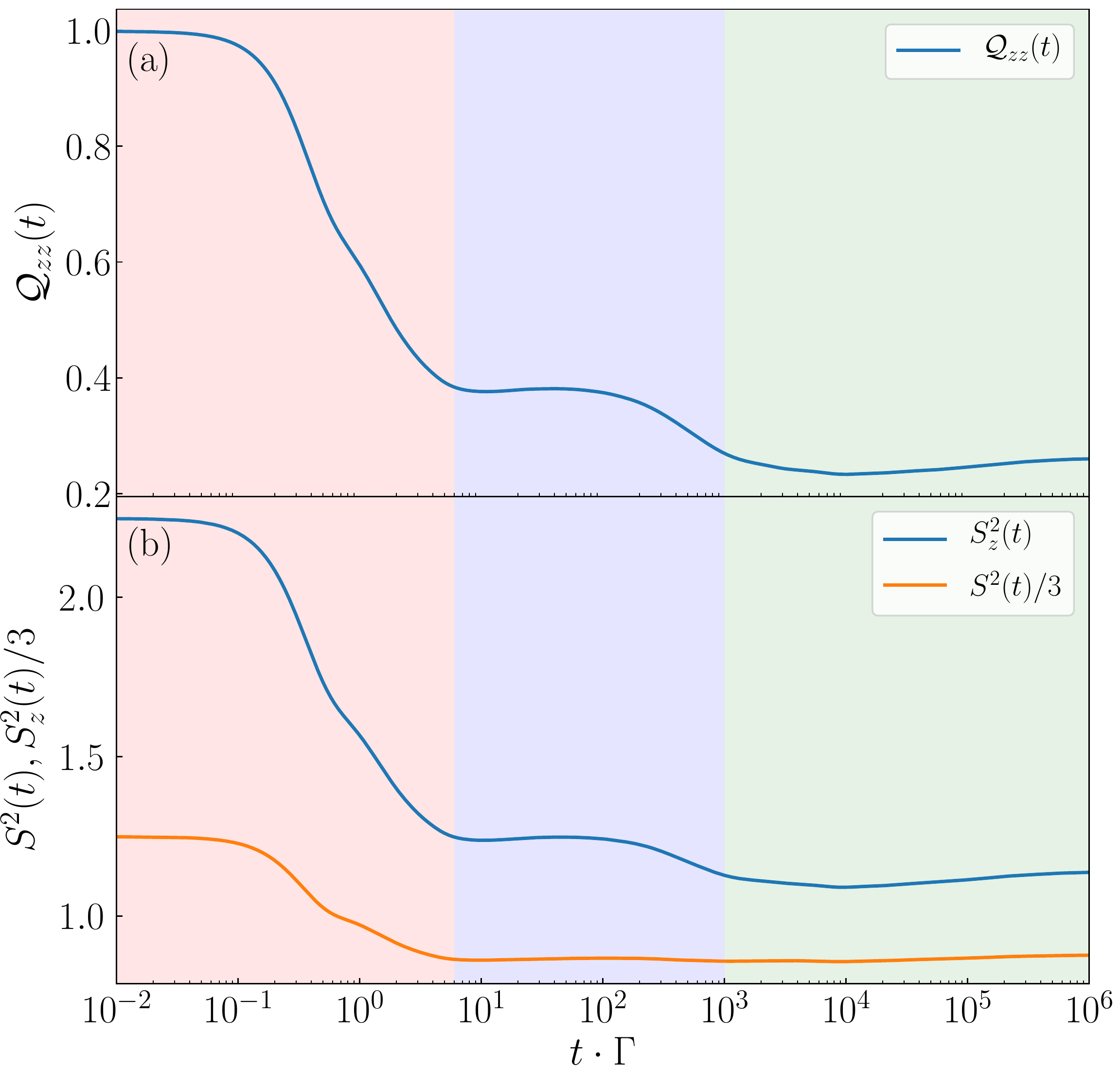}
  \caption{\label{Fig:5}
  The time dependence of the spin-quadrupole moment
  $\mathcal{Q}_{zz}(t)$ and the expectation values of $S^2_z(t)$ and $S^2(t)/3$
  after quench in the coupling strength from $\Gamma_0/U=0.002$ to $\Gamma/U=0.4$.
  The exchange coupling is set to $J/U=0.01$.
  The other parameters are the same as in Fig. \ref{Fig:2}.
}
\end{figure}

In order to better understand the effect of exchange coupling on the spin-quadrupole moment dynamics,
we examine the time-dependence of two spin operators, $S^2_z(t)$ and $S^2(t)/3$,
which are the constituent components of the operator $\mathcal{Q}_{zz}(t)$.
The corresponding plots are shown in Fig. \ref{Fig:5}.
The results were obtained for exchange coupling set to $J/U=0.01$,
where the time dependence reveals two stages of magnetic moment's reduction.

In Fig.~\ref{Fig:5}(a) we display the time-dependent spin-quadrupole moment after the quench in
the coupling strength, while in Fig.~\ref{Fig:5}(b) we present the
time-dependent expectation values of spin operators
$S^2_z(t)$ and $S^2(t)/3$.

To highlight the most important stages of the dynamics,
we marked three time ranges with respective pale colors.
The red background contains short-time dynamics,
when a rapid suppression of all expectation values
takes places in a similar manner like in the case of weaker $J$ coupling.
This dynamics is associated with the change of the ground state
and the reduction of the total spin $S$ from $S=3/2$ to $S=1$
due to the Kondo screening of the orbital spin. This stage of time evolution ends up
at times around $t \cdot \Gamma \approx 10^1$.
Consequently, the next section is marked with purple
background and contains the further dynamics associated with influence of the exchange interaction $J$.
For times $t \cdot \Gamma \gtrsim 10^1$, the time-dependence
of the spin operators clearly shows that the total spin of the system $S$
remains intact and has already achieved the long-time limit.
Further ongoing dynamics takes place exclusively
for $S^2_z(t)$ in the form of second stage of slow suppression of the total
spin-quadrupole moment to $ \mathcal{Q}_{zz}(t) \lesssim 1/3$.
The total time of this stage evolution greatly depends on
the strength of $J$, as shown in Fig. \ref{Fig:4},
i.e. the system faster reaches equilibrium
for higher values of the exchange coupling $J$.

\subsection{Influence of leads' spin polarization}

\begin{figure}[h]
  \includegraphics[width=1\columnwidth]{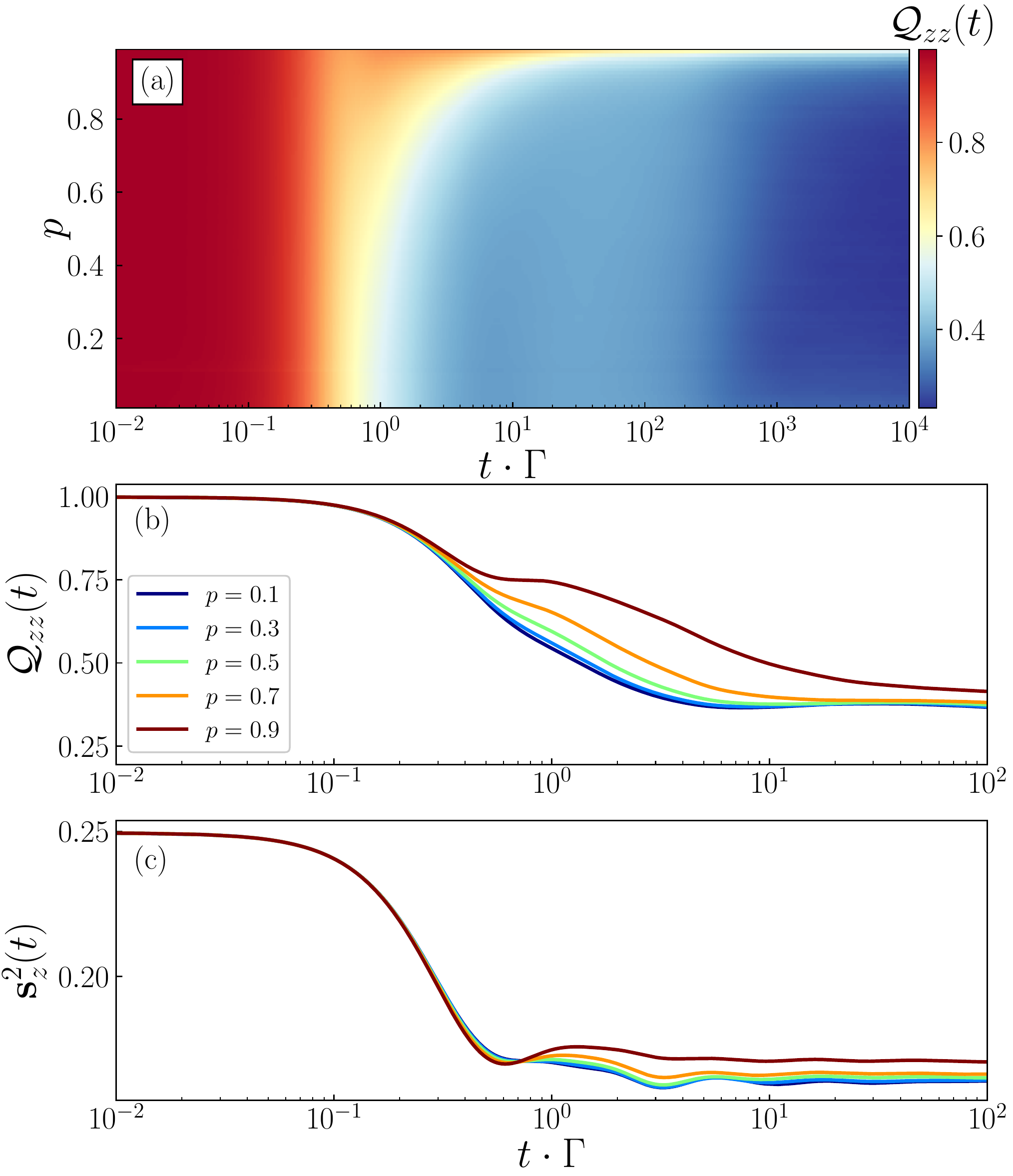}
  \caption{\label{Fig:6}
  (a) The spin-quadrupole moment $\mathcal{Q}_{zz}(t)$
  after a quench from the weak
  to the strong coupling regime
  as a function of time and spin polarization $p$.
  (b) The spin-quadrupole moment $\mathcal{Q}_{zz}(t)$
  and (c) square of the molecular level's magnetization
  $\textbf{s}_{z}^2(t)$
  after the quench as a function of $t \cdot \Gamma$ shown
  for several values of $p$.
  The other parameters are the same as in Fig.~\ref{Fig:5}.
}
\end{figure}

The ferromagnetism of the leads is of great importance for the considered molecular system.
In general, when a finite spin polarization of electrodes is assumed ($p>0$) and their magnetic
moments are aligned, the effective dipolar and quadrupolar exchange fields are generated \cite{martinekPRB05,Misiorny2013Oct}.
The strengths of these fields and their occurrence strongly depend on the degree of spin polarization $p$.
In our analysis, we focus on the quadrupolar term by tuning the molecular level to
the particle-hole symmetry point $\varepsilon=-U/2$. In such a configuration,
the charge fluctuations for both spin directions are equal and independently of the magnitude of $p$
the dipolar field is canceled. Moreover, it was shown that for large-spin molecules, a very small
spin polarization ($p=0.01$) of the leads can give rise to $\langle \mathcal{Q}_{zz} \rangle=S(2S-1)/3=1$.
In equilibrium, a similar dependence to that shown in Fig \ref{Fig:2} is predicted for wide
range of $p$, with finite temperature suppressing the spin-quadrupole moment
in the regime of very weak coupling strengths \cite{Wojcik2019Jul}.
Therefore, let us now discuss the influence of leads spin polarization $p$ on the dynamics of
the spin-quadrupole moment after the quench from the weak
to the strong coupling regime ($\Gamma_0/U=0.002$ and $\Gamma/U = 0.4$).
The results presenting $\mathcal{Q}_{zz}(t)$ for a wide range of spin polarizations $p$
are shown in Fig. \ref{Fig:6}(a).

In the range of small and moderate spin polarizations ($p \lesssim 0.4 $),
the time evolution of spin quadrupole moment has both qualitatively and quantitatively
similar dependence, with two stages of $\mathcal{Q}_{zz}(t)=1$
reduction present, as discussed in previous section for the case of
significant exchange coupling $J/U=0.01$. However, when the spin polarization is
increased further ($p \gtrsim 0.4$), a new step emerges right before the time $t \cdot \Gamma \approx 1$,
when $ \mathcal{Q}_{zz} (t) \approx 3/4$. This step is elongated as
$p$ is increased, reaching times up to $t \cdot \Gamma \approx 10^1$ for $p=0.9$
and eventually not fully relaxing to $ \mathcal{Q}_{zz} (t) = 1/3$,
when $p \rightarrow 1$. This behavior might seem to be counter-intuitive,
as one could expect that with increased spin-polarization of the leads,
the dynamics of the quench should be faster along with rapid suppression
of the spin-quadrupole moment. Here, we predict quite opposite dependence,
as the dynamics is faster and more straightforward when the spin-polarization
of the leads has low-to-moderate values.

The mechanism responsible for the emergence of an additional step in
$\mathcal{Q}_{zz}(t)$ in the case of highly spin-polarized electrodes
is closely related to an enhanced difference between spin-dependent couplings
$\Gamma^\uparrow=\Gamma(1+p)$ and $\Gamma^\downarrow=\Gamma(1- p)$ whilst $p$ is increased.
The bottleneck of the associated dynamics is governed by the coupling to the minority band,
which in result slows down the dynamics of the spin-quadrupole moment.
To clearly analyze the discussed effect, we show
the spin-quadrupole moment $\mathcal{Q}_{zz}(t)$ in Fig. \ref{Fig:6}(b)
and the square of molecular-level magnetization $\textbf{s}_{z}^2(t)$
in Fig.~\ref{Fig:6}(c)
as a functions of $t \cdot \Gamma$. Here, we recall that molecular-level magnetization
is defined as $\textbf{s}_{z}(t)=(n_{\uparrow}(t)-n_{\downarrow}(t))/2$.
The discussed dynamics revealed for highly spin-polarized lead takes place at times  $t \cdot \Gamma \lesssim 1$.
The effect is evident for $p=0.9$, where the dependence of $\mathcal{Q}_{zz}(t)$
exposes a short plateau with $\mathcal{Q}_{zz}(t) \approx 3/4$.
The time-dependence of the local operator $\textbf{s}_{z}^2(t)$ plays an important role here.
As can be seen in the figure, $\textbf{s}_{z}^2(t)$ is a monotonically decreasing function of time
for $p \lesssim 0.7$ in this time regime.
Interestingly, further enhancement of spin-polarization generates
a subtle oscillation in the time-dependence of $\textbf{s}_{z}^2(t)$.
In that time range, $\Gamma^\uparrow=\Gamma(1+p)$ is responsible for a rapid drop
of $\textbf{s}_{z}^2(t)$ due to decrease of the spin-up component.
However, the time evolution is not balanced by the spin-down component due
to a significantly slower dynamics governed by $\Gamma^\downarrow=\Gamma(1-p)$.
As a result, the molecular level has a minimum in $\textbf{s}_{z}^2(t)$,
which is below the long-time limit thermal value,
at time $t \cdot \Gamma \approx 6\cdot 10^{-1}$, see Fig.~\ref{Fig:6}(c).
Subsequently, a small increase takes places leading to eventual relaxation.
This non-monotonic dependence of $\textbf{s}_{z}^2(t)$ is observed as a result of the interplay
between the spin-dependent couplings and, in consequence,
temporarily pauses the reduction of $ \mathcal{Q}_{zz} (t)$.
A similar dynamical behavior was predicted for
the magnetization of a single quantum dot
system coupled to ferromagnetic lead \cite{Wrzesniewski2019Jul}.

\subsection{Influence of magnitude of molecule's spin}

\begin{figure}[h]
	\includegraphics[width=1\columnwidth]{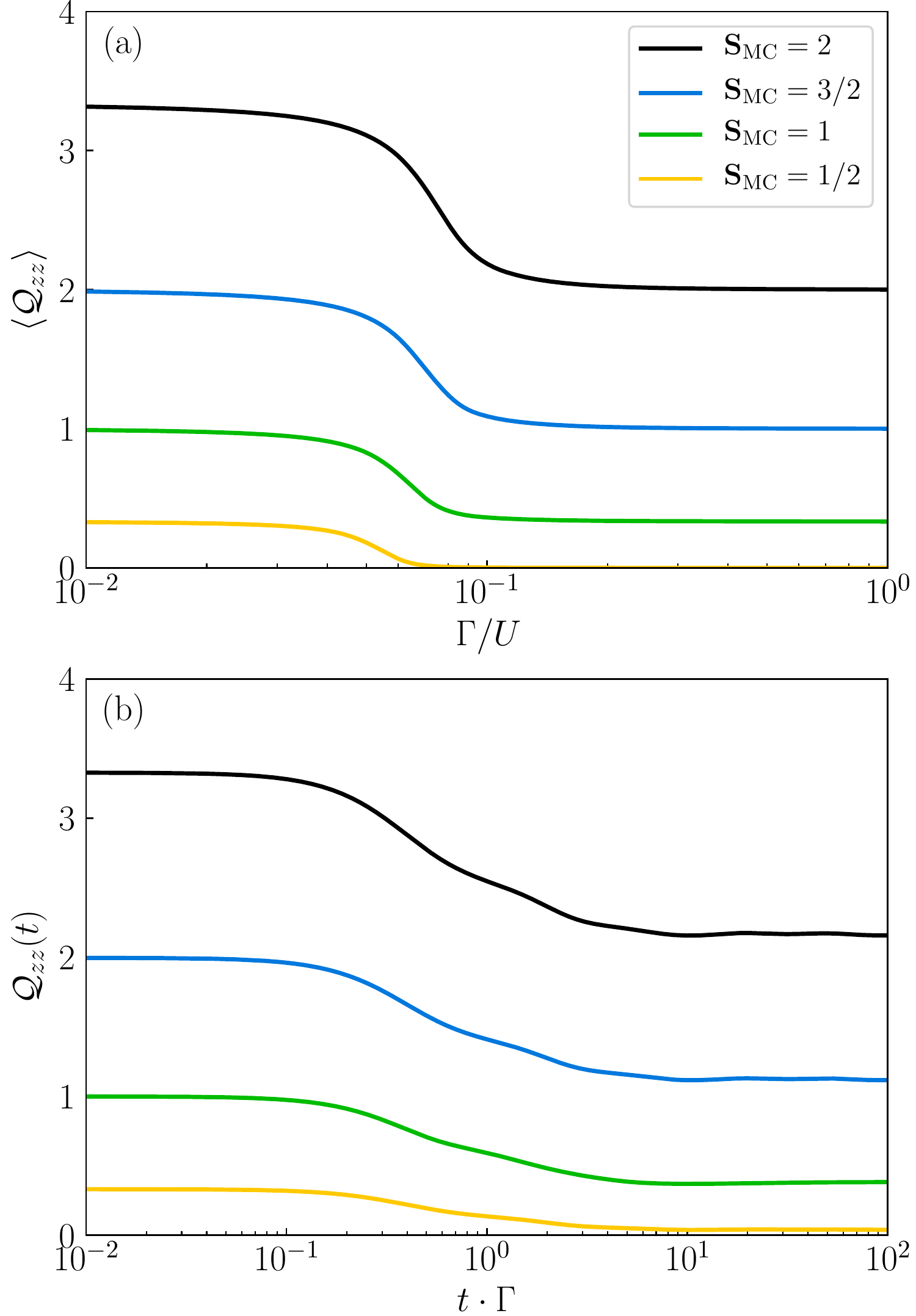}
	\caption{\label{Fig:7}
		(a) The static value of spin-quadrupole moment $\mathcal{Q}_{zz}$ for different values
		of $\textbf{S}_{\rm MC}$ as a function of coupling $\Gamma$ and
		(b) $\mathcal{Q}_{zz}(t)$ for several values
		of $\textbf{S}_{\rm MC}$ as a function of time elapsed after the quench in the coupling $\Gamma$.
		The parameters are the same as in Fig. \ref{Fig:2}.
	}
\end{figure}

In order to generalize our analysis for magnetic molecules of arbitrarily given
total spin $S$, we examine the results of tNRG calculations for models with different
values of magnetic core spin $\textbf{S}_{\rm MC}$.
First of all, we would like to note that in equilibrium, qualitatively
a very similar dependence of $\langle \mathcal{Q}_{zz} \rangle$
on the coupling strength $\Gamma$ is expected independently of the value of $\textbf{S}_{\rm MC}$,
see Fig. \ref{Fig:7}(a). As the spin of magnetic core is increased,
the maximal value of the spin-quadrupole moment in the weak coupling regime
is increasing accordingly with $\langle \mathcal{Q}_{zz} \rangle = S_z^2-S(S+1)/3$.
Furthermore, the transition range of coupling strengths preceding the strong coupling regime
is similar for all cases, i.e. $ 5 \cdot 10^{-2}  \lesssim \Gamma/U  \lesssim 2 \cdot 10^{-1}$.
Eventually, the important change influenced by the total spin of the molecule is in the
value, by which the spin-quadrupole moment is reduced, when the coupling regime is switched
from the weak to the strong one. This difference is enhanced, when the total spin number is increased.

The time-dependent spin-quadrupole moment after the quench as a function of time
for several values of $S_{\rm MC}$ is shown in Fig. \ref{Fig:7}(b).
Here, one can see that the dynamics is still governed by the strength of coupling to electrodes,
as the reduction of spin-quadrupolar moment starts and achieves
long-time limit at similar moments on the time axis rescaled with $\Gamma$.
For all considered values of $\textbf{S}_{\rm MC}$, time-evolutions behave
in a similar fashion with the main distinction of initial and
final values of $\langle \mathcal{Q}_{zz} \rangle$ and the rate of reduction.
Similar effects and dependencies due to the exchange interaction or spin-polarization of the leads
as discussed in previous sections are predicted also for molecules
with even higher total spin number (not shown here).
Therefore, our dynamical studies presented in this work
have a general character and are valid for broad range of SMM systems,
in which the quadrupolar exchange field emerges from ferromagnetic proximity effect.

\subsection{Finite temperature effects}

\begin{figure}[t]
  \includegraphics[width=1\columnwidth]{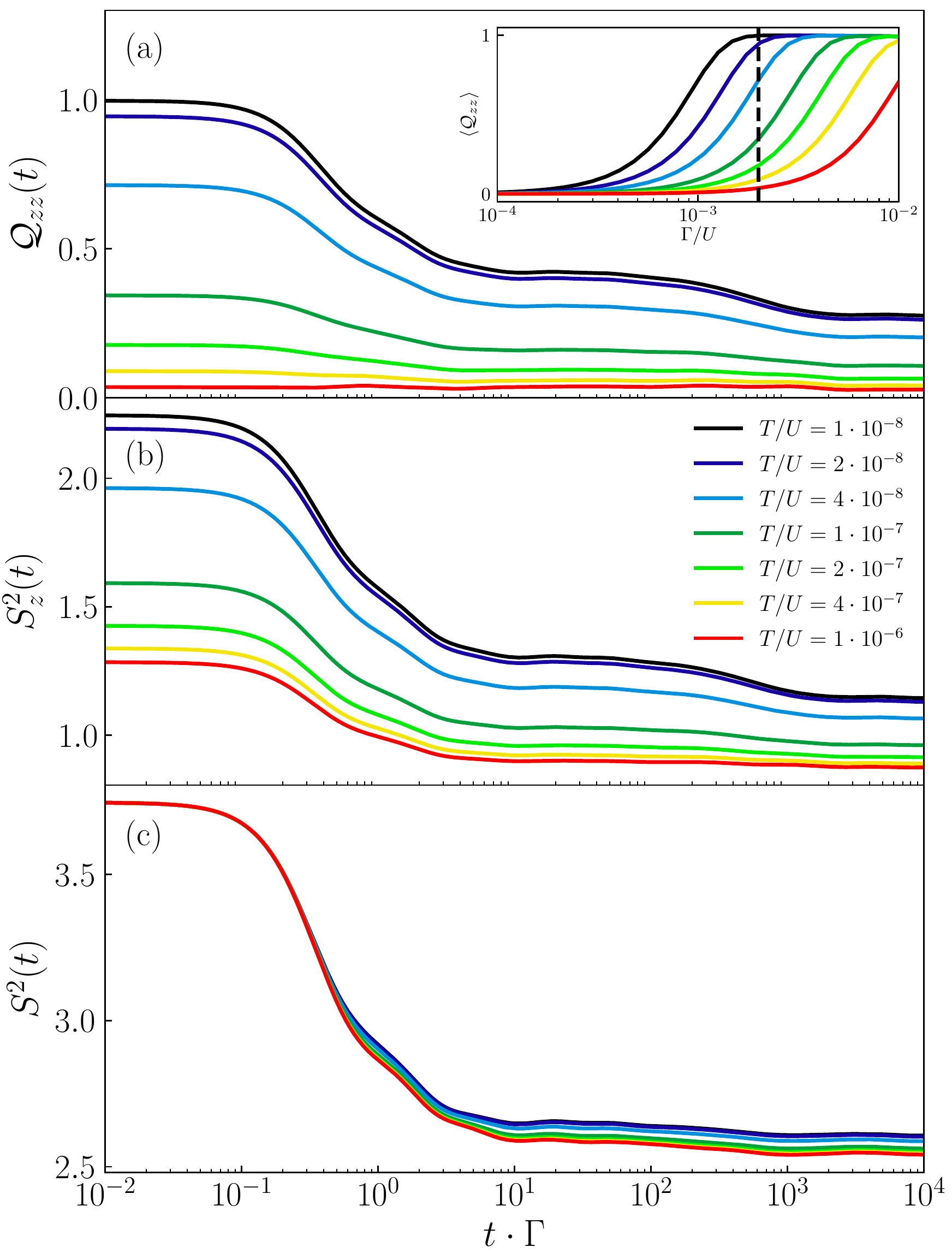}
  \caption{\label{Fig:8}
  (a) The spin-quadrupole moment $\mathcal{Q}_{zz}(t)$,
  (b) $S_z^2(t)$ and (c) $S^2(t)$
  for several values of temperature $T$ plotted as a function of time elapsed after the quench in $\Gamma$.
  The inset in (a) shows the influence of temperature on the static value
  of the spin-quadrupole moment in the weak coupling regime.
  Dashed vertical line indicates the initial value of the coupling strength.
  The parameters are the same as in Fig. \ref{Fig:5}.
}
\end{figure}

Lastly, we examine the influence of finite temperature $T$ on the
spin-quadrupole moment and its time evolution following the quench in the coupling.
In Fig. \ref{Fig:8} we show the time-dependent expectation value of the spin-quadrupole moment as
well as the corresponding spin operators for several values of temperatures.

For the regime of very low temperatures, $T/U \lesssim 10^{-8}$,
both the initial and final values, as well as the quench dynamics
of the spin-quadrupole moment remain similar to the case of zero temperature.
Further increase of temperature, however, has a significant impact
on the system's behavior. First of all, the initial value of the
spin quadrupole moment in the weak coupling regime is significantly
suppressed as the temperature is increased. The inset in
Fig. \ref{Fig:8}(a) shows how the increase of temperature reduces
the static spin-quadrupole moment in the weak coupling regime.
The vertical dashed line represents the initial coupling strength
$\Gamma_0/U=0.002$, we have used in the evaluation of the quench dynamics.
From the above considerations, it is evident that the fine tuning of the
coupling strength for the initial state is critical,
in particular when the suppression of the maximal value of the moment is expected.
Furthermore, the long-time limit value of the time-dependent spin-quadrupole moment
is also reduced by the temperature, however, not as strongly as the initial value.
In consequence, as the temperature is increased, the whole transient dynamics is
exposing more moderate dependence, with almost completely flat one
for temperatures $T/U \gtrsim 10 ^{-6}$.
It is also important to note, that all the characteristic time scales
discussed in earlier analysis for $T=0$ are conserved for the finite temperatures,
where a considerable suppression of spin-quadrupole moment is predicted.
In particular, the rapid reduction associated with the change of the ground state
that dominates the dynamics for times $t \cdot \Gamma \lesssim 10^1$ and
further slower dynamics influenced by the exchange coupling taking place
for times up to $t \cdot \Gamma \approx 10^3$ are all still noticeable for
temperatures up to $T/U \approx 10^{-7}$.

The inspection of the spin operators $S_z^2(t)$ and $S^2(t)$,
see Figs. \ref{Fig:8}(b) and \ref{Fig:8}(c), brings the conclusion
that the temperature influences spin-quadrupole moment
only by reduction of the $z$-th component of the SMM's spin.
Meanwhile the square value of the total spin remains unaffected,
as it corresponds to the change of the molecule's ground state.
This fact clearly indicates the reduction of the molecule's magnetic anisotropy
due to thermal fluctuations.


\section{Conclusions} \label{conclusions}

We have analyzed the quench dynamics of large-spin
magnetic molecules attached to spin-polarized ferromagnetic leads.
The study was performed by using the
time-dependent numerical renormalization group method.
We focused on the dynamics of the spin-quadrupole moment
and a quench associated with switching the system from the weak coupling regime
to the strong coupling one, in which the Kondo correlations are present
and the screening of the molecular level spin develops.

In general, we have shown that the time necessary to achieve the new thermal value
in the strong coupling regime is inversely proportional to squared coupling strength,
i.e. $t \propto 1/\Gamma^2$. Furthermore, we examined the role of ferromagnetic
exchange coupling $J$ and showed that when the magnitude of this interaction is considerable,
additional step in the time-dependence emerges, corresponding to slower
dynamics mediated by $J$, which is an interesting case
of interplay between the Kondo correlations and ferromagnetism exposed
in our dynamical studies. The influence of electrodes' spin-polarization
was also discussed, strongly indicating that the quench dynamics
is faster and more straightforward for low-to-average values.
We also generalized our studies to systems with higher total spin
numbers to show that observations and conclusions are valid for a wide class
of SMM systems with arbitrary total spin number.

Finally, we studied the influence of finite temperature on the
spin-quadrupole moment. A strong suppression of the time-dependent
value is predicted in certain range of temperatures, due to
reduction of the molecule's anisotropy. The important role
of the coupling strength fine tuning is also indicated,
which may have relevance for the experiments.


\begin{acknowledgments}
This work was supported by the Polish National Science
Centre from funds awarded through the decision No. 2017/27/B/ST3/00621.
The computing time at the Pozna\'n Supercomputing and Networking Center is acknowledged.
\end{acknowledgments}


%

\end{document}